\documentclass[%
reprint,
superscriptaddress,
amsmath,amssymb,
aps,
prb,
]{revtex4-2}

\usepackage{graphicx}
\usepackage{dcolumn}
\usepackage{bm}
\usepackage[colorlinks=true,allcolors=blue]{hyperref}

\begin{document}

\preprint{APS/123-QED}

\title{Interdependent scaling of long-range oxygen and magnetic ordering in non-stoichiometric Nd${}_2$NiO${}_{4.10}$}

\author{Sumit Ranjan Maity}
\affiliation{Laboratory for Neutron Scattering and Imaging, Paul Scherrer Institut, Villigen CH-5232, Switzerland}

\affiliation{University of Geneva, Department of Quantum Matter Physics (DQMP) 24, Quai Ernest Ansermet CH-1211 Gen\`{e}ve 4, Switzerland}

\author{Monica Ceretti}
\affiliation{ICGM, Université de Montpellier, CNRS, ENSCM, FR-34095 Montpellier, France}

\author{Lukas Keller}
\affiliation{Laboratory for Neutron Scattering and Imaging, Paul Scherrer Institut, Villigen CH-5232, Switzerland}

\author{J\"{u}rg Schefer}
\affiliation{Laboratory for Neutron Scattering and Imaging, Paul Scherrer Institut, Villigen CH-5232, Switzerland}

\author{Martin Meven}
\affiliation{Institute of Crystallography, RWTH Aachen University and Jülich Centre for Neutron Science (JCNS),
	Forschungszentrum Jülich GmbH at Heinz Maier-Leibnitz Zentrum (MLZ), DE-857478 Garching, Germany}

\author{Ekaterina Pomjakushina}

\affiliation{Laboratory for Multiscale Materials Experiments, Paul Scherrer Institut, Villigen CH-5232, Switzerland}

\author{Werner Paulus}
\affiliation{ICGM, Université de Montpellier, CNRS, ENSCM, FR-34095 Montpellier, France}

\date{\today}

\begin{abstract}
	
	Hole doping in Nd${}_{2}$NiO${}_{4.00}$ can be either achieved by substituting the trivalent Nd atoms by bivalent alkaline earth metals or by oxygen doping, yielding Nd${}_{2}$NiO${}_{4+\delta}$. While the alkaline earth metal atoms are statistically distributed on the rare-earth sites, the extra oxygen atoms in interstitial lattice remain mobile down to ambient temperature and allow complex ordering scenarios depending on $\delta$ and T. Thereby the oxygen ordering, usually setting in far above room temperature, adds an additional degree of freedom on top of charge, spin and orbital ordering, which appear at much lower temperatures. In this study, we investigated the interplay between oxygen and spin ordering for a low oxygen doping concentration i.e. Nd${}_{2}$NiO${}_{4.10}$. Although the extra oxygen doping level remains rather modest with only one out of 20 possible interstitial tetrahedral lattice sites occupied, we observed by single crystal neutron diffraction the presence of a complex 3D modulated structure related to oxygen ordering already at ambient, the modulation vectors being $\pm$2/13\textit{\textbf{a*}}$\pm$3/13\textit{\textbf{b*}}, $\pm$3/13\textit{\textbf{b*}}$\pm$2/13\textit{\textbf{b*}} and $\pm$1/5\textit{\textbf{a*}}$\pm$1/2\textit{\textbf{c*}} and satellite reflections up to fourth order. Temperature dependent neutron diffraction studies indicate the coexistence of oxygen and magnetic ordering below T${}_{N}$ $\simeq$ 48 K, the wave vector of the Ni sublattice being \textbf{\textit{k}}=(100). In addition, magnetic satellite reflections adapt exactly the same modulation vectors as found for the oxygen ordering, evidencing a unique coexistence of 3D modulated ordering for spin and oxygen ordering in Nd${}_{2}$NiO${}_{4.10}$. Temperature dependent measurements of magnetic intensities suggest two magnetic phase transitions below 48 K and 20 K, indicating two distinct onsets of magnetic ordering for the Ni and Nd sublattice, respectively.  

\end{abstract}
 
\maketitle


\section{Introduction}

\begin{figure}

\includegraphics[scale=0.3]{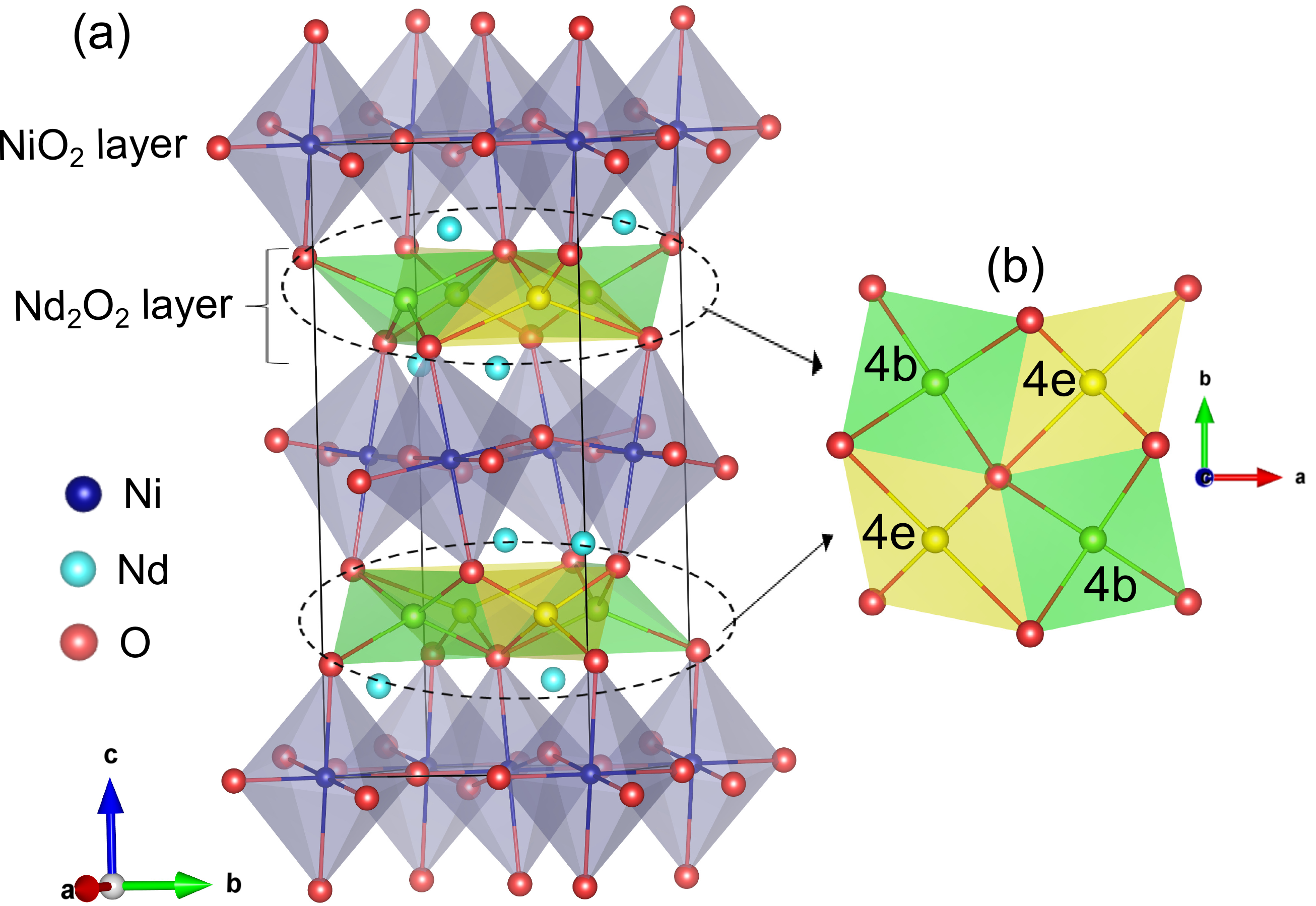}

	\caption{\label{fig:crystal structure} The proposed LTT Crystal structure of Nd${}_{2}$NiO${}_{4.10}$ at room temperature (S.G. P4${}_{2}$/\textit{ncm}). (a) The tetragonal unit cell consists of alternate stacking of NiO${}_{2}$ and Nd${}_{2}$O${}_{2}$ layers along the \textit{c}-axis. (b) Two different tetrahedral sites with Wyckoff symmetry 4\textit{b} and 4\textit{e}. Excess oxygen atoms selectively occupy only the 4$b$ Wyckoff sites as observed with powder and single crystal neutron diffraction studies.}
	
\end{figure}

	Perovskite-type layered rare earth Ruddlesden-Popper nickelates with the chemical formula Ln${}_{2}$NiO${}_{4}$ (Ln: La, Pr, Nd) are one of the many strongly correlated electronic systems which show interesting physical and electronic properties as a function of temperature \cite{Demourgues1996,Freeman2004,Freeman2002,Hiroi1990,Hucker2004,Minervini2000,Bassat2013a,Perrichon2015}. These materials are complex because of the different active degrees of freedom such as lattice, spin, charge and orbital ordering which interact in a competitive and synergetic way and open a wide field for applications. The crystal structures of these compounds can be described as a layered structure with alternate stacking of NiO${}_{2 }$ layers and rock-salt type Ln${}_{2}$O${}_{2}$ bilayers along the long axis (\textit{c}-axis) of the unit cell (\textit{cf.} Fig.\ref{fig:crystal structure}) \cite{Jorgensen1989}. The undoped compounds are Mott insulators. Interesting physical properties like a metal-insulator transition appear when nickelates are doped with excess charge carriers. Holes can be incorporated either by partial substitution of Ln${}^{3+}$ cations with divalent cations like Sr${}^{2+}$, Ca${}^{2+}$ and Ba${}^{2+}$ or by doping with excess oxygen atoms. However, a metal-insulator transition occurs at a relatively large hole concentration ($n{}_{h}$) of 1 hole/formula unit \cite{Tranquada1996a}. In the insulating state, holes segregate inside the NiO${}_{2}$ planes as an ordered array of parallel regions separating the antiferromagnetic (AFM) spin order as a domain boundary, known as the stripe/checkerboard charge ordered state \cite{Tranquada1996a,Tranquada1995,Kajimoto2003,Yoshizawa2000a,Freeman2004,Carlson2004}. The incorporated holes are in a diluted region for $n{}_{h} < $ 0.25 and strong correlations among the holes appear for 0.25 $\leq n{}_{h} \leq$ 1 that consequently results in stripe/checkerboard charge ordered states, which are experimentally observed with neutron diffraction in Sr-and O-doped Ln${}_{2}$NiO${}_{4}$. 

	\noindent The substitution of Ln${}^{3+}$ ions with smaller divalent cations is random and has a relatively weak effect on the lattice. In contrast, for oxygen doped systems, excess oxygen atoms take interstitial lattice sites inside the Ln${}_{2}$O${}_{2}$ bi-layer and introduce local disorder in the long-range tilting pattern of NiO${}_{6}$ octahedra by systematically pushing adjacent apical oxygen atoms towards the nearest vacant interstitial sites \cite{Wochner1998,Ceretti2015,Ceretti2018,Maity2019}. Such apical oxygen displacements are strongly anharmonic in character and are important for the oxygen mobility close to room temperature, resembling to a successively in- and exhaling behavior of the involved empty and occupied tetrahedra along the diffusion pathway \cite{Ceretti2015,Maity2019}. Neutron and synchrotron diffraction experiments have shown evidence of long-range ordering of excess oxygen atoms in the interstitial sites \cite{Tranquada1995a,Ceretti2015,Hiroi1990,LeDreau2012,Demourgues1993a,Demourgues1993}. The presence of oxygen superstructure reflections in La${}_{2}$NiO${}_{4+\delta}$ with 0.05 $\mathrm{\le}$ $\delta$ $\mathrm{\le}$ 0.11 have been explained by a 1D staging model in which intercalated layers of excess oxygen atoms cluster periodically spaced along the $c$-axis \cite{Tranquada1994}. A stage-$n$ order consists of single excess oxygen atoms layer separated by $n$ NiO${}_{2 }$ planes along the $c$-axis. The Coulomb repulsion between ionic excess oxygen atoms limit the density within an interstitial layer, and also the spacing between intercalated layers along the $c$-axis. Consequently, with decreasing $\delta$ from 0.105 to 0.06, the spacing between intercalated layers increases, and the value of $n$ increases from 2 to 4 \cite{Tranquada1994}. Stronger correlations among the excess oxygen atoms appear with increasing excess oxygen doping concentration $\delta$. Consequently, a 3D ordering pattern of excess oxygen atoms is evidenced at $\delta$ $\simeq$ 0.125 \cite{Tranquada1994a}. Very recently, a complex 3D ordering of excess oxygen atoms has been reported for homologous Pr${}_{2}$NiO${}_{4+\delta}$ with $\delta$ $\simeq$ 0.22-0.25 i.e. with a $\delta$ value close to the most non-stoichiometric limit revealing translational periodicities of almost 100 {\AA} \cite{Dutta2020}. The ordered arrangement of excess oxygen atoms leads to an ordered deformation of NiO${}_{2}$ planes, which provides a modulated potential that can pin the charge stripes to the lattice. Therefore, the amount of excess oxygen atoms and its ordering state play a crucial role on the nature of spin and charge ordering appearing in these compounds at low temperatures. More intriguingly, the magnetic ordering in La${}_{2}$NiO${}_{4+\delta}$ at $\delta$ $>$ 0.110 changes from commensurate to incommensurate stripe order and the charge ordering appears with a change of correlations among excess oxygen atoms, from 1D staging to 3D long-range order \cite{Tranquada1994a,Zhang2005}. It was proposed that the charge correlations fluctuate about an average ordered configuration determined by the ordering of excess oxygen atoms and a direct relationship between the modulation vectors of the charge order and the 3D interstitial order has been observed along one direction of the lattice \cite{Wochner1998}.

	\noindent The pinning of charge and spin correlations by the structural distortions and anomalous suppression of the superconductivity has been observed in the low temperature tetragonal (LTT) phase of La${}_{1.48}$Nd${}_{0.4}$Sr${}_{0.12}$CuO${}_{4}$ \cite{Tranquada1996c} with P4${}_{2}$/\textit{ncm} space group. Very recently, a similar LTT phase has been reported for the moderately oxygen doped (Pr/Nd)${}_{2}$NiO${}_{4+\delta}$ compounds with \textit{$\delta$} $\sim$ 0.125 \cite{Maity2019,Ceretti2018}. Surprisingly, no long-range ordering of the excess oxygen atoms has been observed down to 2 K with powder diffraction studies.  In addition, a commensurate magnetic structure has been proposed for Nd${}_{2}$NiO${}_{4.11}$ with propagation vector \textbf{\textit{k}}=(100) \cite{Maity2019}. The commensurate magnetic order coexists with an additional magnetic phase which has been assumed to be incommensurate but significantly different from the stripe spin ordering \cite{Maity2019}. 
	
	\noindent In this work, we aim to understand this additional magnetic phase by employing neutron diffraction studies on a Nd${}_{2}$NiO${}_{4+\delta}$ single crystal with similar oxygen non-stoichiometry of about $\delta$ $\simeq$ 0.10. In addition, we search for possible superstructure reflections appearing from the ordering of excess oxygen atoms, spin and charge in the LTT phase of Nd${}_{2}$NiO${}_{4.10}$

\section{experimental methods}

\subsection{Sample synthesis}

	A large single-crystal was grown at Institute Charles Gerhardt Montpellier (ICGM) using a two-mirror optical floating zone furnace at high temperature in oxygen atmosphere as described elsewhere \cite{Wahyudi2015,Maity2020a}. The as-grown crystal has an overall oxygen non-stoichiometry ($\delta$) in the range of 0.23-0.25. Consequently, further heat treatment was performed to obtain a single crystal with reduced excess oxygen content $\delta$. For that, a small section ($\sim$ 200 mg) was cut and annealed in a dynamic vacuum at 1073 K for 6 h. The details are reported in reference \cite{Maity2019}. The overall oxygen stoichiometry in the single crystal was determined to be 4.097(10) with the least-squares refinement of the room temperature single crystal neutron diffraction data \cite{Maity2019}. A powder sample was separately synthesized and reduced under identical conditions used for the single crystal. The overall oxygen stoichiometry in the powder sample was determined using thermogravimetric analysis (TGA) at high temperature as well as with neutron powder diffraction data at 300 K to be 0.11(2), and thus very similar to the one used for the single crystal using neutron diffraction experiments. Both the single crystal and powder samples are thus identical to those reported in reference \cite{Maity2019}. In between different measurements, the samples were stored inside a glove box with Ar atmosphere to minimize the oxidizing effects.
	
\subsection{X-ray and neutron diffraction measurements}	

	\noindent Synchrotron X-ray powder diffraction (XRPD) measurements were performed at different temperatures in between 5-300 K at the Material Sciences (MS) beamline X04SA at the Swiss light source (SLS) of the Paul Scherrer Institute (PSI), Switzerland. The wavelength ($\lambda$ = 0.6208(1) \AA) and instrumental resolution parameters were determined from a standard LaB${}_{6}$ powder (NIST), measured under identical experimental conditions. Samples were filled into thin glass capillaries of 0.3 mm diameter and continuously rotated to reduce preferred orientation during the data collection. Neutron powder diffraction (NPD) studies were carried out in the temperature range of 2-300 K using the cold neutron powder diffractometer DMC \cite{Schefer1990, DMC} at SINQ, PSI in Switzerland. Both X-ray and neutron diffraction data were analyzed with the FullProf suite program \cite{Rodriguez-Carvajal1993}. DMC was further used to perform the reciprocal plane mappings on the single crystal at variable temperatures between 2 and 300 K. A pre-oriented single crystal sample was mounted on top of a thin aluminium pin and the reciprocal (${hk}$0) and (${h}$0$l$) planes were mapped out by rotating the single crystal around the respective zone axis. The reciprocal maps are particularly useful to locate weak superstructure peaks or diffuse scattering in the reciprocal space with either structural and/or magnetic origin. For all the measurements, a wavelength of 2.4586(3) {\AA} was used to access a reciprocal space up to a momentum transfer [Q=4$\pi$sin$\theta$/$\lambda$] of about 4.15 {\AA}${}^{-1}$. Moreover, $\lambda$/2 contamination could be excluded at this wavelength on DMC. The reciprocal planes were reconstructed from the raw data using TVtueb program \cite{tvnexus}. Integrated intensities of selected structural and magnetic Bragg reflections were measured using the hot-neutron four-circle diffractometer HEiDi (equipped with a point detector) at the Heinz Maier-Leibnitz Zentrum (MLZ) at the FRM-II reactor in Garching, Germany \cite{Meven2015}. For these measurements, transverse i.e. $\omega$-scans were performed with the crystal oriented on the ($hk$0) scattering plane with the $c$-axis vertically aligned. A short wavelength of 0.793(1) {\AA} was used for the data collection up to a momentum transfer of about 11.30 {\AA}${}^{-1}$. The data sets were collected at 300 K, 55 K and 2 K. Structural least-squares refinements of the integrated and Lorentz corrected structure factors of P-type Bragg reflections were performed with the JANA2006 software package \cite{Petricek2014}. The crystal structure is visualized using VESTA freeware \cite{Momma2011}. 
	
\section{Results and discussions}

\subsection{\label{sec:level2} Average crystal structure}

\begin{figure}

	\includegraphics[scale=0.2]{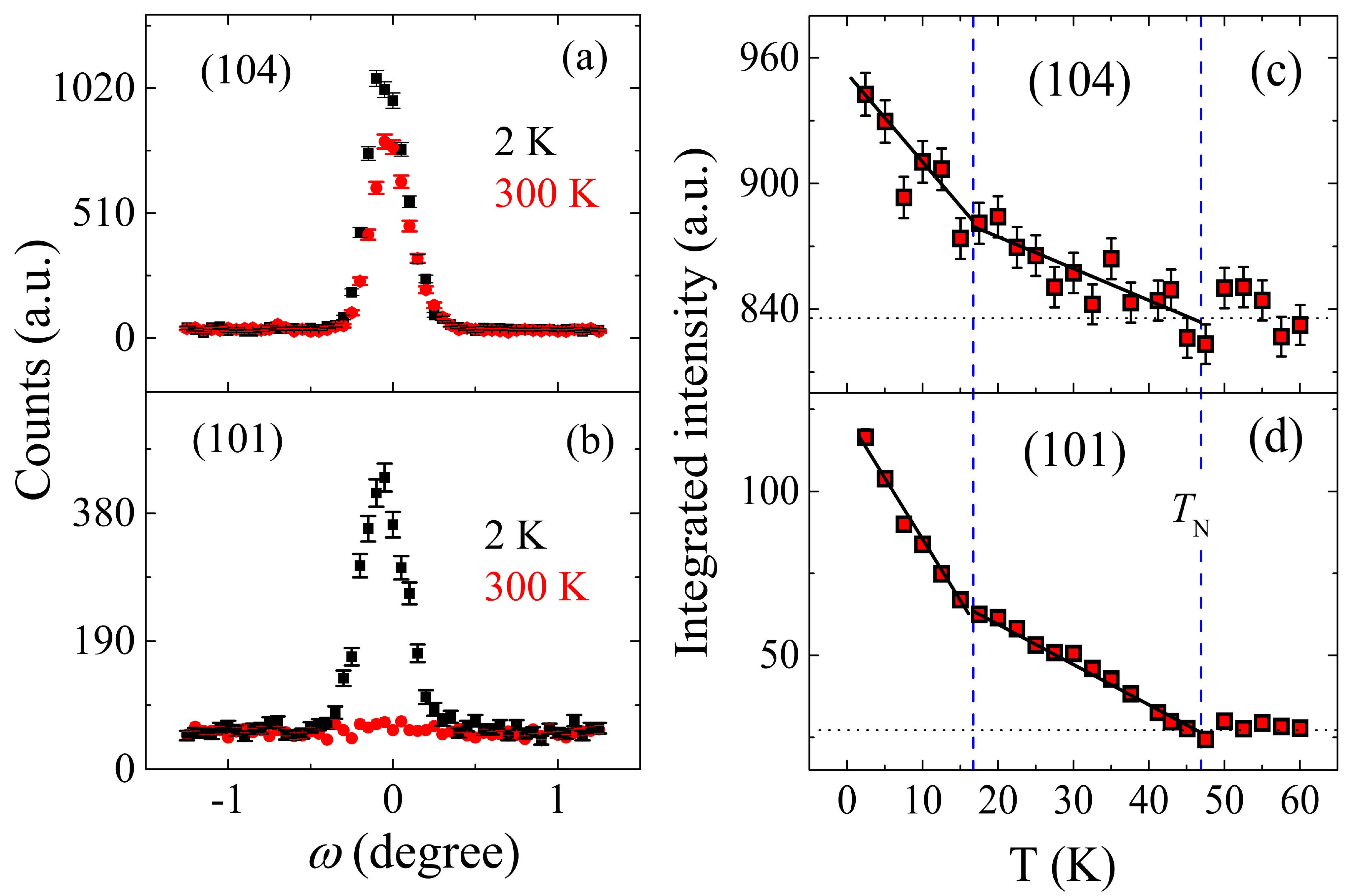}

	\caption{\label{fig:omega scans} The $\omega$-scan profiles ($\omega$ values around $\theta$${}_{hkl}$) of (a) (104) and (b) (101) Bragg peaks recorded at different temperatures on a Nd${}_{2}$NiO${}_{4.10}$ single crystal. Temperature evolution of the integrated intensities of (c) (104) and (d) (101) Bragg peaks. Vertical blue dashed lines at T$\sim$48 K and T$\sim$18 K indicate the magnetic ordering temperature of the Ni-sublattice and the onset of magnetic polarization of the Nd-sublattice, respectively. Thick solid lines are only to guide the eye. The measurements were carried out on HEiDi, MLZ with $\lambda$=0.793(1) {\AA}.}

\end{figure}

	The LTT crystal structure of the moderately oxygen doped Nd${}_{2}$NiO${}_{4+\delta}$ compound with $\delta$ $\sim$ 0.11 has been previously revealed with powder and single crystal neutron diffraction studies in the temperature range of 2-300 K \cite{Maity2019}. The unit cell of the LTT phase is shown in Fig. \ref{fig:crystal structure}(a) obtained from room temperature neutron diffraction data. In general, the Nd${}_{2}$O${}_{2}$ layer is in the extensional strain whereas the NiO${}_{2}$ layer feels a compressional strain. The structural strain is released through structural distortions that involve a systematic tilting of NiO${}_{6}$ octahedra about a fixed crystallographic axis. In the LTT phase, NiO${}_{6}$ octahedra are canted around the [1$\overline{1}$0] direction in the basal plane such that in-plane nearest octahedra are contra-rotated. The rotation axis alternates between the [110] and [1$\overline{1}$0] direction in the neighboring NiO${}_{2}$ planes along the \textit{c}-axis. Such an octahedra tilting gives rise to extra P-type superstructure reflections compared to the fully disordered high temperature tetragonal (HTT) phase with space group F4/\textit{mmm}. Our neutron diffraction measurements give evidence of sharp P-type superstructure reflections already at room temperature indicating the LTT phase. The $\omega$-scan profile of such P-type structural Bragg peaks measured at room temperature is presented in Fig. \ref{fig:omega scans}(a) for the (104) reflection. The specific octahedra tilting in the LTT phase allows favourable openings for the tetrahedral 4\textit{b} (3/4, 1/4, 1/4) and 4\textit{e} (1/4, 1/4, $z\sim$ 1/4) interstitial sites as shown in Fig. \ref{fig:crystal structure}(b). The refinement of the integrated intensity data of the main structural Bragg peaks at room temperature demonstrates that the excess oxygen atoms selectively occupy the 4\textit{b} interstitial sites, in accordance with previously reported results \cite{Maity2019,Ceretti2018}. It is worth to note that the LTT structure has been reported for the stoichiometric Nd${}_{2}$NiO${}_{4.00}$ compound below 135 K \cite{Rodriguez-Carvajal1990a}. Here, the LTT structure becomes stable already at ambient temperature related to additional oxygen doping. We note that only one of 20 possible tetrahedral sites is occupied in Nd${}_{2}$NiO${}_{4.10}$ i.e. 0.4 oxygen atoms are present on average per unit cell. 

\begin{figure*}
	
	\includegraphics[scale=0.39]{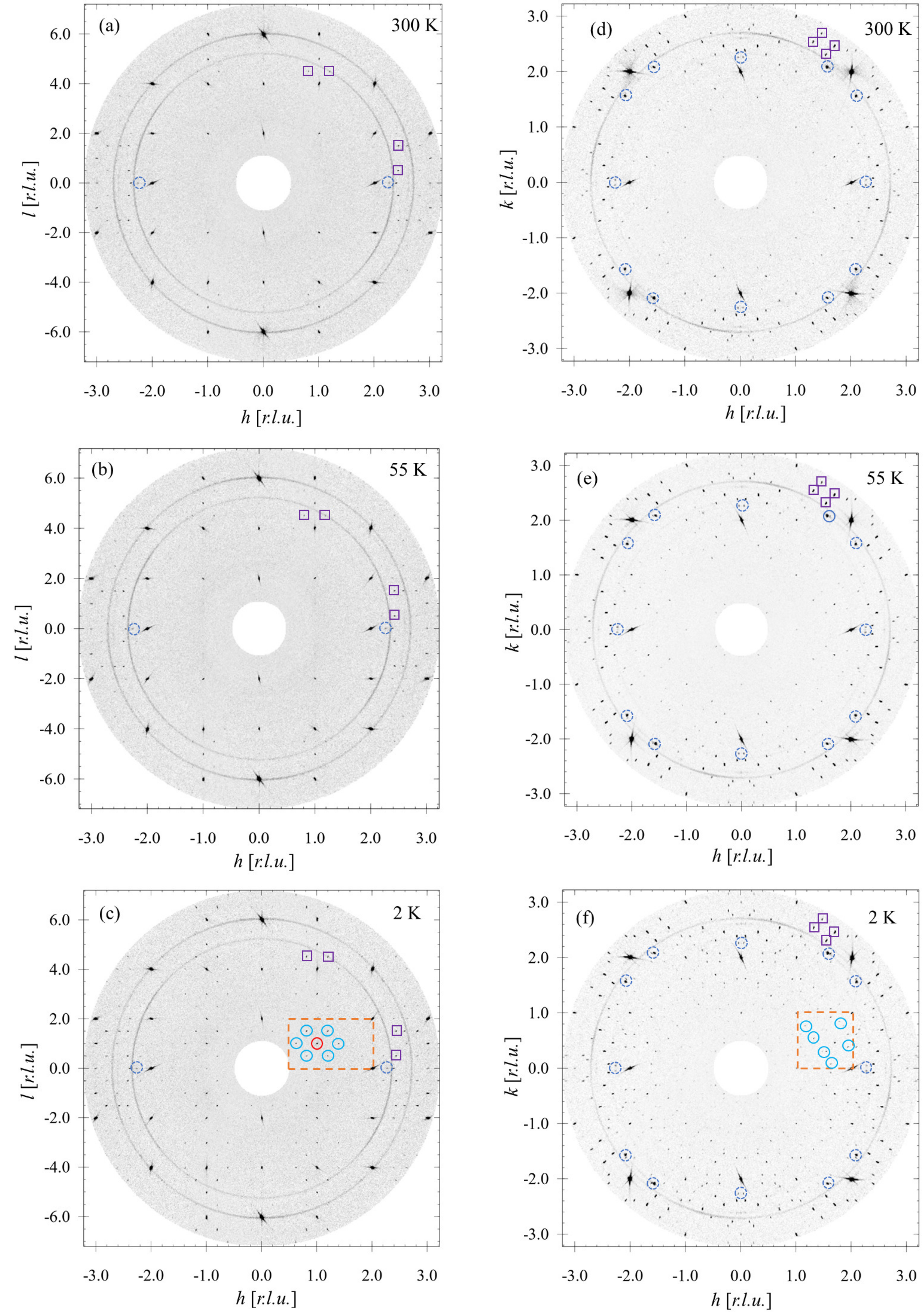}
	
	\caption{\label{fig:reciprocal plane} Reconstructed single crystal neutron diffraction plane maps of Nd${}_{2}$NiO${}_{4.10}$. Excerpts of (a)-(c) ($h$0$l$) and (d)-(f) ($hk$0) reciprocal planes measured at 300 K, 55 K and 2 K. The dashed blue circles show the reflections emerged from the NiO impurity phase while the squares and full circles represent characteristic superstructure reflections appearing from the ordering of excess oxygen atoms and spins in Nd${}_{2}$NiO${}_{4.10}$, respectively. Rectangular boxes in orange in the 2 K maps display the sections which are magnified in Fig. \ref{fig:cuts}. The measurements were carried out on DMC at SINQ, PSI with $\lambda$=2.4586(3) {\AA}.}
	
\end{figure*}

     \begin{figure*}
     	
     	\includegraphics[scale=0.22]{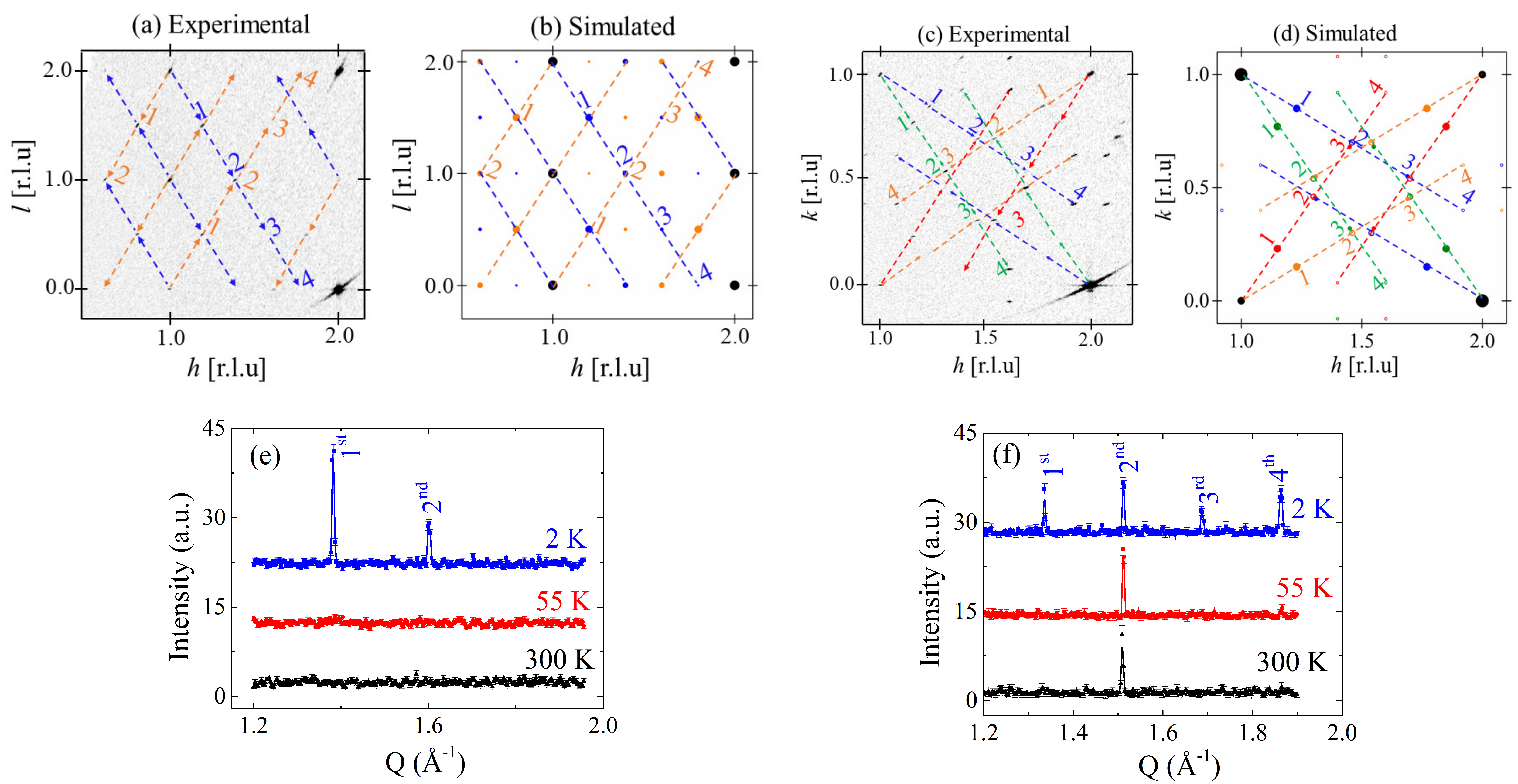}
     	
     	\caption{\label{fig:cuts} Magnified zones of the (a) experimentally measured ($h$0$l$) plane presented together with the (b) simulated zone. Magnified sections of the (c) experimentally measured ($hk$0) plane presented along with the (d) simulated zone. Integer numbers represent different order of the satellites corresponding to a Bragg peak. Q-scans through the superstructure peaks observed in the (${hk}$0) and (${h}$0${l}$) plane maps of Nd${}_{2}$NiO${}_{4.10}$. The Q-scan in (e) is starting from (102) along the blue line up to the 4th order satellite as shown in (a) and in (f), the cut is starting from (110) along the green line up to the 4th order satellite as shown in (c). A constant y-offset is used to plot the 55 K and 2 K data.}  
     	
     \end{figure*}

\subsection{Commensurate magnetic structure}

	The DC magnetic susceptibility data, recorded for the Nd${}_{2}$NiO${}_{4.10}$ single crystal with a field of 1T applied parallel to an arbitrary crystallographic axis in zero-field-cooled (ZFC) configuration, is presented in Fig. S1 of the supplemental material. An antiferromagnetic transition at T${}_{N}$ $\simeq$ 48 K can be clearly discerned from the magnetic susceptibility plot. The magnetic structure of Nd${}_{2}$NiO${}_{4.10}$ below T${}_{N}$ can be described with a commensurate propagation vector \textbf{\textit{k}}=(100) as previously revealed by NPD data. Consequently, the intensities of all Bragg peaks significantly increase below T${}_{N}$ due to the magnetic contribution and commensurate magnetic peaks become evident with decreasing temperature from 55 K to 2 K. The $\omega$-scan profile of a pure magnetic Bragg peak such as the (101) is presented in Fig. \ref{fig:omega scans}(b). The temperature evolution of integrated intensities of the (104) and (101) reflections are shown in Fig. \ref{fig:omega scans}(c) and \ref{fig:omega scans}(d), respectively. Two magnetic transitions can clearly be distinguished from the temperature dependence. The first starts below T $\simeq$ 48 K in accordance with the DC magnetic susceptibility data and indicates the onset of antiferromagnetic order of the Ni-sublattice. The one indicated by a change of the slope below T $\simeq$ 20 K, while intensities of all the Bragg reflections are rapidly increasing, associated to the onset of magnetic polarization of the Nd-sublattice. The commensurate magnetic structure of the average structure corresponding to LTT phase has been solved by the least-squares refinement from a set of integrated intensities recorded at 2 K. In this magnetic structure, Ni and Nd spins are coupled antiferromagnetically and oriented parallel to the crystallographic [110] or [1$\overline{1}$0] direction with a weak ferromagnetic component along the \textit{c}-axis due to the presence of strong Dzyaloshinskii–Moriya interaction (DMI). The total magnetic moments for the Ni and Nd-sites were refined to be 1.02(5) \textit{$\mu$}${}_{B}$ and 1.08(2) \textit{$\mu$}${}_{B}$ at 2 K, respectively, in good agreement with previous results \cite{Maity2019}.

\subsection{Excess oxygen ordering and the additional magnetic phase}%

	 Neutron diffraction maps of the (${h}$0${l}$) and (${hk}$0) planes are presented in Fig. \ref{fig:reciprocal plane}(a)-(f) for 300 K, 55 K and 2 K. In addition to the main structural Bragg peaks, sharp superstructure reflections are also clearly observed, of which a few are already present at ambient temperature (\textit{cf}. Fig. \ref{fig:reciprocal plane}(a) and (d)). Their intensities are significant and reach up to 1-2\% with respect to the main structural Bragg spots, signifying the presence of complex structural modulations already at 300 K. Moreover, these reflections are strongest at higher Q ($\simeq$ 2.5-3.8 \AA$^{-1}$) and presumably due to the long-range ordering of excess oxygen atoms and related structural distortions. This indicates that excess oxygen atoms in Nd${}_{2}$NiO${}_{4.10}$ are not distributed randomly in the interstitial sites. Consequently, a complex 3D ordering of excess oxygen atoms gets established even with only one out of 20 possible tetrahedral sites is occupied (i.e. 0.4 oxygen atoms on average per P-type unit cell). In this regard, a similar LTT phase is observed for oxygen intercalated La${}_{2}$NiO${}_{4+\delta}$ in the range of 0.02 $\lesssim \delta \lesssim$ 0.03 for which long-range ordering of the excess oxygen atoms was detected at any temperatures \cite{Tranquada1994,Rice1993}. In addition, 1D staging of the excess oxygen atoms along the $c$-axis was reported for La${}_{2}$NiO${}_{4+\delta}$ with similar $\delta$ concentration \cite{Tranquada1994}. Thus, compared to La${}_{2}$NiO${}_{4+\delta}$, the observed 3D ordering of excess oxygen atoms in Nd${}_{2}$NiO${}_{4.10}$ is quite surprising. We suppose that the observed long-range ordering of excess oxygen atoms is partially due to the lower ionic radius of Nd$^{3+}$ ions (1.163 {\AA} in nine-fold coordination) compared to the La$^{3+}$ ions (1.216 {\AA} in nine-fold coordination). The reciprocal plane maps remain almost unchanged with decreasing temperature down to 55 K, thus signifying the existence of a similar ordering pattern of excess oxygen atoms at 55 K. However, new and strong superstructure reflections in addition to the commensurate magnetic Bragg peaks appear below T${}_{N}$ which are clearly evident in the 2 K maps (\textit{cf}. Fig. \ref{fig:reciprocal plane}(c) and (f)). As these new superlattice peaks rise only below T${}_{N}$ and are quite strong at low Q ($\simeq$ 0.37-1.88 \AA$^{-1}$), they are most likely of magnetic origin, in agreement with our magnetic susceptibility measurements. Therefore, the commensurate magnetic order coexists with a hidden magnetic phase in oxygen doped Nd${}_{2}$NiO${}_{4.10}$ as previously pointed out from NPD studies \cite{Maity2019}. These magnetic superstructure reflections can be rationalized with the same modulation vectors that define the excess oxygen ordering reflections at 300 K. For moderately oxygen doped Nd${}_{2}$NiO${}_{4.10}$, a commensurate magnetic order coexists along with long-range oxygen ordering, which was not previously observed in homologous La${}_{2}$NiO${}_{4+\delta}$ compounds. Furthermore, extra commensurate reflections such as the (210) and the (105) which are symmetry forbidden in P4${}_{2}$/\textit{ncm}, were evidenced already at ambient temperature, suggesting that the true symmetry of Nd${}_{2}$NiO${}_{4.10}$ is possibly lower while inconsistent with the structural model proposed for homologous La${}_{2}$NiO${}_{4.105}$ \cite{Tranquada1993}. In this context, it is worth to note that the commensurate charge ordering reflections in hole doped La${}_{2-x}$Sr${}_{x}$NiO${}_{4}$ are found to emerge in such commensurate positions \cite{Yoshizawa2000a}. However, we do not expect stripe or checkerboard type charge orders to occur in our sample as the required hole concentration for such correlation to appear is considerably higher than the incorporated hole concentration in our sample being only about 0.2 per formula unit. Furthermore, in the presence of checkerboard type charge ordering, magnetic Bragg peaks at ($h \sim$ 0.5, 0, 0) positions would also be expected, which are clearly not observed here \cite{Yoshizawa2000a}. Therefore, the commensurate reflections are most likely due to a lowering of the symmetry from P4${}_{2}$/\textit{ncm}.
	 	
	\begin{figure*}
		
		\includegraphics[scale=0.30]{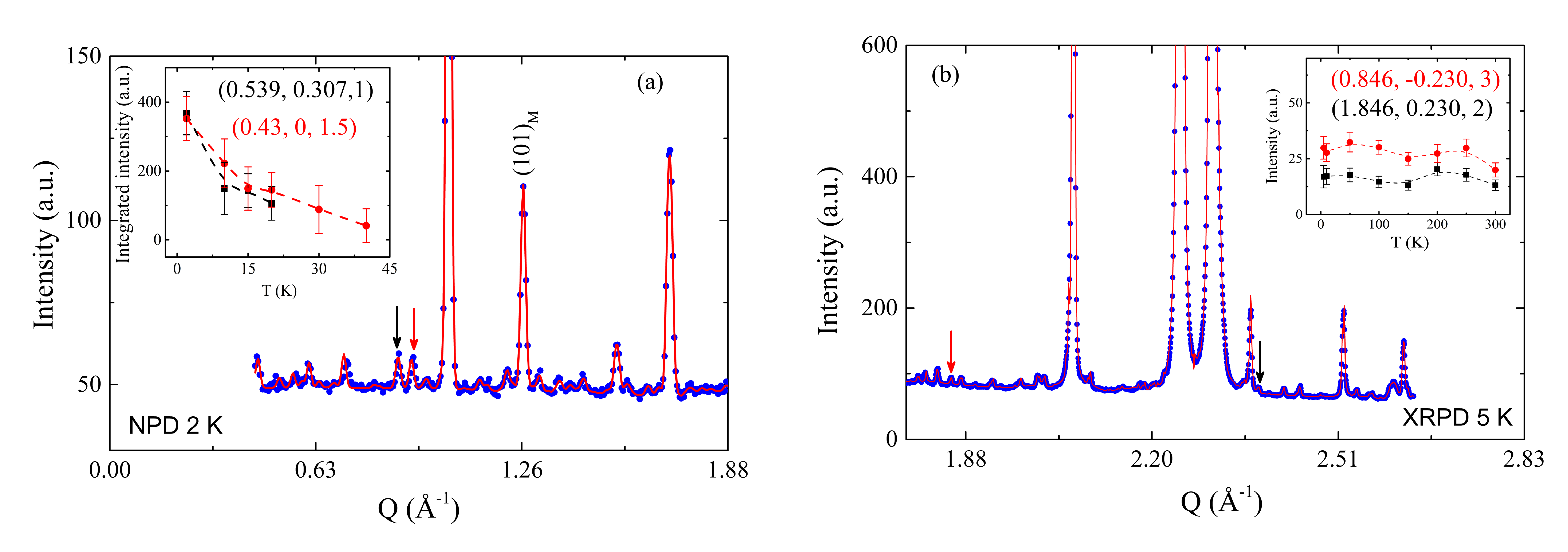}
		
		\caption{\label{fig:powder data} Powder diffraction patterns of Nd${}_{2}$NiO${}_{4.11}$. Excerpts of the Le Bail fit of the (a) NPD data at 2K and (b) XRPD data at 5 K with the $q$-vectors obtained from the reciprocal space plane mappings. Observed (blue circles) and calculated (red line) patterns resulting from the profile analysis of the powder data. Both insets represent the temperature dependence of two superstructure reflections as marked by colored arrows in the powder patterns. The NPD data was measured on DMC at SINQ, PSI with $\lambda$=2.4586(3) {\AA}. The XRPD data was measured on MS-X04SA at SLS, PSI with $\lambda$=0.62104(3) {\AA}. Both powder data are identical to those reported in reference \cite{Maity2019}.}
		
	\end{figure*} 
	 
 	\noindent We have attempted to index all the superstructure reflections appearing in the 2 K maps. This is a complicated task especially in case the presence of a large number of satellites i.e. higher order harmonics. It is found that all the superstructure reflections could be indexed assuming modulation vectors of type \textbf{$q{}_{n}$} = $\alpha{}_{n}$\textit{\textbf{a*}}+$\beta{}_{n}$\textit{\textbf{b*}}+$\gamma{}_{n}$\textit{\textbf{c*}} added to the position of each Bragg reflection \textit{\textbf{G}}=$h$\textit{\textbf{a*}}+$k$\textit{\textbf{b*}}+$l$\textit{\textbf{c*}}+$m$\textbf{$q{}_{n}$}, with $m$ being an integer number representing the order of corresponding superstructure reflections. The underlying modulation vectors, which describe both the excess oxygen ordering and magnetic ordering in the ($hk$0) and ($h$0$l$) planes, are found to be $\pm$2/13\textit{\textbf{a*}}$\pm$3/13\textit{\textbf{b*}}, $\pm$3/13\textit{\textbf{a*}}$\pm$2/13\textit{\textbf{b*}} and $\pm$1/5\textit{\textbf{a*}}$\pm$1/2\textit{\textbf{c*}}. In addition, structural Bragg peaks from epitaxially intergrown NiO as reported in \cite{Dutta2020} were also observed at all the investigated temperatures. An example of the manually indexed experimental and idealized ($hk$0) plane is presented in Fig. S2 and Fig. S3 of the supplemental material, respectively. The simulated reconstructions for possible superstructure reflections positions were undertaken using these modulation vectors. However, it is essential to consider all the symmetry-related directions of the modulation vectors to satisfactorily describe all the satellite positions in the ($hk$0) and ($h$0$l$) plane maps. This is clearly evident from the ($hk$0) plane map in which all the satellites positions at 2 K are excellently represented by the all eight directions of the modulation vectors are given by $\pm$2/13\textit{\textbf{a*}}$\pm$3/13\textit{\textbf{b*}} and $\pm$3/13\textit{\textbf{a*}}$\pm$2/13\textit{\textbf{b*}}. In a similar way, the ($h$0$l$) plane can be described by the four directions of the modulation vectors which are characterized by $\pm$1/5\textit{\textbf{a*}}$\pm$1/2\textit{\textbf{c*}}. It is furthermore necessary to consider up to 4th order harmonics for a successful indexation of all the satellites appearing in both reciprocal planes at 2 K. The simulated reconstructions, which excellently represent the experimental maps, are shown in Figs. S4 and S5 of the supplemental material. The magnified sections of the experimental and simulated zones of the ($h$0$l$) and ($hk$0) planes are displayed in Fig. \ref{fig:cuts} (a)-(d). It is found that the superstructure peaks in the (${h}$0${l}$) plane are symmetrically positioned in such a way that they form a hexagon around a commensurate magnetic peak of (${h}$0${l}$) type with $l$ = odd as shown in Fig. \ref{fig:cuts} (b). An important finding of these simulations is that a substantial overlap of intensities from different harmonics associated to different Bragg peaks occurs in the reciprocal plane. For example, in the ($h$0$l$) plane (\textit{cf}. Fig. \ref{fig:cuts} (b)), the first and second order satellites related to the 1/5\textit{\textbf{a*}}+1/2\textit{\textbf{c*}} vector starting from (101) Bragg peak are identical with the first and second order satellites corresponding to the 1/5\textit{\textbf{a*}}-1/2\textit{\textbf{c*}} vector starting from (102) and (100) Bragg peaks, respectively. However, in the ($hk$0) plane (\textit{cf}. Fig. \ref{fig:cuts} (d)), the satellites are sufficiently separated which allows to determine their structure factors satellites separately. Figs. \ref{fig:cuts}(d) and \ref{fig:cuts}(e) represent specific Q-scans through the superstructure reflections in the reciprocal maps obtained at different temperatures. It is evident that the intensities of some superstructure reflections like the first and second harmonics related to the (102) and all the harmonics associated to (110) are strongly enhanced below T${}_{N}$ owing to strong magnetic contributions. Further, the modulation vectors are invariant in the temperature range of 2-300 K i.e. the same modulation vectors satisfactorily define the excess oxygen ordering peaks at room temperature. Moreover, no significant change in the full widths at half maximum (FWHMs) of the oxygen superstructure reflections is detected as a function of temperature. 

	\noindent The excerpts of XRPD and NPD data recorded at 5 K and 2 K are shown in Figs. \ref{fig:powder data}(a) and \ref{fig:powder data}(b), respectively. We note that superstructure reflections are observed in the NPD data only below T${}_{N}$ as shown in Fig. S6 of the supplemental material. In strong contrast, superstructure reflections are present in the XRPD data already at room temperature and their positions remain unchanged down to 5 K. The satellites in the XRPD data are very weak owing to the small X-ray form factor of oxygen atoms but are clearly distinguishable from the background. The integrated intensities of two characteristic superstructure reflections observed in the NPD and XRPD data are plotted in the insets of Figs. \ref{fig:powder data}(a) and \ref{fig:powder data}(b). It can be seen that their integrated intensities in the NPD data are strongly enhanced with decreasing temperature down to 2 K while in the XRPD data remain invariant in the temperature range of 2-300 K. Therefore, the superlattice peaks in the XRPD data are essentially due to the ordering of excess oxygen atoms whereas those observed in the NPD data below T${}_{N}$ can be attributed to the magnetic ordering. Even though neutrons are quite sensitive to light oxygen atoms, the superstructure reflections originating from the excess oxygen ordering appear in the reciprocal plane maps only at higher Q but are not observed with NPD data probably because of their superimposition with the main Bragg reflections (see Fig. S7). Finally, we have performed Le Bail fits of the XRPD and the NPD data in order to index the superstructure reflections using the modulation vectors which are obtained from the plane maps. All the superstructure peaks observed by X-ray and neutron powder diffraction can be indexed satisfactorily by these modulation vectors as shown in Figs. \ref{fig:powder data}(a) and \ref{fig:powder data}(b). The fully refined NPD pattern is presented in Fig. S7 of the supplemental material. This result indicates that the modulation vectors obtained from the plane mappings are equally valid for the description of satellites observed in the powder data and further confirms that the magnetic order is correlated to the excess oxygen order in moderately oxygen doped Nd${}_{2}$NiO${}_{4+\delta}$ with $\delta\simeq$ 0.10.

\section{Conclusions}

	Large scale oxygen and magnetic ordering have been evidenced by single crystal neutron diffraction studies on Nd${}_{2}$NiO${}_{4.10}$ to reveal identical modulation vectors resulting in a 3D modulated structure. The commensurate antiferromagnetic order in Nd${}_{2}$NiO${}_{4.10}$ sets in at T${}_{N}$ $\simeq$ 48 K with a propagation vector \textbf{\textit{k}}=(100). The most peculiar point here is that the magnetic ordering modulation vectors are no longer independent and exactly follow the modulation vectors of the oxygen ordering. This is evidenced unambiguously from the selected planes explored with single crystal neutron diffraction and also from the refinement of powder neutron diffraction data. The magnetic order thus coexists below T${}_{N}$ with the 3D ordering of excess oxygen atoms in Nd${}_{2}$NiO${}_{4.10}$, which is atypical so far, and not previously encountered for the homologous nickelates. Nd${}_{2}$NiO${}_{4.10}$ thus seems to present a special case in terms of structural and magnetic ordering phenomena, compared to other nickelates and especially to the La${}_{2}$NiO${}_{4.10}$. We note that no evidence for charge ordering has been identified in this study. Although the oxygen stoichiometry remains quite diluted with only one out of 20 possible tetrahedral sites occupied (i.e. 0.4 oxygen atoms on average per P-type unit cell), a complex 3D ordering gets established. This clearly indicates the potential of the interstitial oxygen atoms to enable inter- and intra-layer coupling for structural and magnetic ordering. As a phenomenon, this becomes important with respect to the recently evidenced sub-mesoscale oxygen ordering in oxygen rich Pr${}_{2}$NiO${}_{4+\delta}$($\delta$ $\simeq$ 0.25), revealing translational periodicities of almost 100 \AA. In this context, the mediation achieved between magnetic and oxygen ordering, thus yields a promising concept to establish long-range magnetic ordering, essentially based on long-range oxygen ordering.
	
	\noindent Oxygen-doped Ruddlesden Popper type oxides are entirely different compared to their Sr-doped homologues. The latter require high reaction temperatures for doping resulting into a statistically distribution of immobile Sr$^{2+}$ ions, while excess oxygen atoms remain mobile down to ambient temperature, allowing to establish different superstructures related to ordering of the interstitial oxygen atoms as a function of temperature and doping concentration. In this context, it would be interesting to explore the structural and electronic response especially for lower oxygen doping concentrations, with the aim to probe the limits of possible correlations linking oxygen with magnetic and charge ordering. In that context, it would be of importance to know how the degree of freedom for electronic ordering phenomena will scale with the oxygen ordering.
	
\begin{acknowledgments}
	
	This work is based on experiments performed at Paul Scherrer Institute (PSI), Villigen, Switzerland and at the Heinz Maier-Leibnitz Zentrum (MLZ), Garching, Germany. The authors acknowledge the beam times used on DMC at PSI/SINQ and on HEiDi at MLZ/FRM II (operated jointly by RWTH Aachen University and FZ J\"{u}lich within JARA collaboration) as well as laboratory equipment and support from LDM/PSI and the ``Plateforme d'Analyse et de Caract\`{e}risation'' of the ICG Montpellier. The authors gratefully acknowledge the financial support from the Swiss National Science Foundation (SNF) through grant 200021L\_157131 and the French National Research Agency (ANR) through grant 14-CE36-0006-01 of the SECTOR project.

\end{acknowledgments}


%


\end{document}